\documentclass{article}
\usepackage{emulateapj}
\usepackage{apjfonts}
\usepackage{timesfonts}
\usepackage{psfig}

\begin{document}

\def\theta{\vartheta}
\def\phi{\varphi}
\def\sax{{\it Beppo}SAX}

\title{On the absorption feature in the prompt X--ray spectrum of GRB~990705}

\author{Davide Lazzati}
\affil{Institute of Astronomy, University of Cambridge,
Madingley Road, Cambridge CB3 0HA,
England \\ e--mail: lazzati@ast.cam.ac.uk}
\author{Gabriele Ghisellini}
\affil{Osservatorio Astronomico di Brera, Via E. Bianchi 46 I--23807 
Merate, Italy}
\author{Lorenzo Amati and Filippo Frontera\altaffilmark{1}}
\affil{Istituto TESRE, CNR, Area della Ricerca di Bologna, Via Gobetti 101,
I--40129 Bologna, Italy}
\author{Mario Vietri}
\affil{Universit\`a di Roma 3, Via della Vasca Navale 84, I--00147 Roma, 
Italy}
\and\author{Luigi Stella}
\affil{Osservatorio Astronomico di Roma, Via Frascati 33, I--00040 Monteporzio 
Catone, Italy}
\altaffiltext{1}{Also: Universit\`a di Ferrara, Dipartimento di Fisica, 
via Paradiso 12, I--44100 Ferrara, Italy}

\begin{abstract}
The absorption feature detected in the prompt X--ray emission of 
GRB~990705 bears important consequences. 
We investigate different production mechanisms and
we conclude that the absorbing material cannot be very 
close to the burster and is likely to be moderately clumped.
These properties challenge any model in which the burst explodes
in coincidence with the core--collapse of a massive rotating star.
We show that the straightforward interpretation of the absorption
feature as a photoionization K edge of neutral iron faces a
severe problem in that it requires a huge amount of iron in the 
close vicinity of the burster. 
We then discuss an alternative scenario, in which iron ions
are kept in a high ionization state by the burst flux, 
and the absorption feature is produced by resonant scattering from 
hydrogen--like iron, broadened by a range outflow velocities.
In this case the physical conditions and geometry of the absorbing material 
are fully consistent with the presence of a young supernova remnant  
surrounding the burst site at a radius $R \sim 10^{16}$ cm.
We finally discuss how this remnant might affect the generation of 
afterglows with a standard power--law flux decay.
\end{abstract}


\section{Introduction}

Emission or absorption features in the X--ray spectrum of gamma ray bursts 
(GRBs) and their afterglows provide a fundamental tool to study their close 
environment and thus their possible progenitors.
To date, four bursts showed evidence for an iron emission line during the 
X--ray afterglow, observed 8--40 hours after the burst event (GRB 970508, 
Piro et al., 1999; GRB 970828, Yoshida et al., 1999; GRB 991216, Piro et 
al., 2000; GRB 000214, Antonelli et al., 2000). Here we concentrate instead 
on GRB 990705, which showed a prominent absorption feature at 3.8 keV and 
an equivalent hydrogen column density, that disappeared 13 s after the burst 
onset (Amati et al., 2000, hereafter A2000).
This burst was observed with the Gamma Ray Burst Monitor (GRBM) and the Wide
Field Cameras (WFC) of \sax. It had a duration of $\sim$42 s in the GRBM
and $\sim 60$ s in the WFC. In the entire WFC-GRBM band
(2--700 keV) the fluence was $(9.3 \pm 0.2) \times 10^{-5}$ erg cm$^{-2}$
(A2000). The absorption feature was interpreted by A2000 as due to an
edge produced by neutral iron (i.e. at 7.1 keV),
redshifted to 3.8$\pm$0.3 keV (corresponding to $z=0.86\pm0.17$).
Optical observations of the host galaxy give a redshift
compatible with what inferred from the X--rays: $z_{\rm opt} = 0.84$
(Andersen et al., 2001). Fitting the spectrum with an absorbed 
($N_H = (3.5 \pm 1.4) \times 10^{22}$~cm$^{-2}$) power law plus an 
absorption edge yielded $\tau =1.4\pm 0.4$.
Most important for our analysis, the absorption feature,
present in the first 13 seconds from the trigger, become undetectable 
afterwords. During its entire duration, GRB~990705 showed very pronounced
short--term variability with flux variations $> 50\%$ on timescales
of a fraction of a second.

The prompt emission was followed by a fading X--ray afterglow, detected
by the Narrow Field Instruments of \sax~11 hours after
the trigger, with a 2--10 keV flux of $(1.9\pm 0.6)\times 10^{-13}$ 
erg s$^{-1}$ cm$^{-2}$ (A2000).
An infrared ($H\sim 16.7$, 0.28 days after the GRB) and faint optical 
transient ($V\sim 22.5$, 0.73 days after the GRB) were
discovered by Masetti et al. (2000), while no radio afterglow was detected
(Subrahmanyan et al., 1999; Hurley et al., 1999).

In principle, an absorption feature can be produced by a reflecting 
slab, even when the illuminating continuum is directly 
visible to the observer. 
But in the case of fast and extremely variable events such as GRBs
this can be excluded on the basis of the following arguments.
First, the GRB continuum is produced by a highly relativistic fireball, 
which becomes thin to Thomson scattering at radii $\gtrsim 10^{13}$ cm.
Therefore any reflector should have a size at least comparable to 
this radius. Moreover it would produce uncollimated radiation: the 
reflected flux would therefore be
reduced with respect to the direct continuum reaching the observer.
Any feature (in absorption or emission) would then be greatly diluted. 
Moreover, the light travel path of the direct continuum photons is
shorter than that of the reflected photons, which then reach the
observer at later times. Therefore we conclude that the absorption feature 
observed in GRB 990705 is due to material located in the line of sight 
between the observer and the fireball.

We investigate here the formation of a prompt absorption feature 
such as that reported for GRB 990705. 
While different scenarios have been proposed to explain the line in emission
(see Lazzati et al., 1999; Vietri et al.,  2001; Rees \& Meszaros, 2000; 
Weth et al., 2000; B\"ottcher, 2000), the properties of the absorption 
feature of GRB 990705 strongly points to a unique scenario, 
in which a few solar masses of iron rich matter are located 
close to the burst explosion site. 
The transient nature of the feature, moreover, places tight constraints
on the absorbing matter--burst distance (we will show that it must lie
between $10^{16}$ and $10^{18}$~cm), and provides useful information on
the geometry of the absorbing material. If future observations will confirm 
the presence of absorption features of similar 
characteristics, then models invoking a simultaneous explosion of the 
burst and the progenitor star would be in serious trouble.
In fact, large amounts of iron at such distances along the line of sight 
can be provided only by a supernova (SN) exploding several months 
before the burst.

\section{General scenarios}
\label{sec:gg}
We here qualitatively discuss the general scenarios
which are investigated in detail in the following sections.

As mentioned in the introduction, we can exclude that the
absorption feature is produced by reflection.
Therefore we will hereinafter assume that there is 
some absorbing material between us and the burst. There are, in principle,
different possibilities.

\vskip 0.3 true cm

\noindent
{\it Edge due to a molecular cloud ---} 
Assume that the feature is produced by a molecular cloud in the host galaxy
of the burst, possibly containing the burst itself, even if this latter
assumption is not mandatory.
In this case the absorption material will have a nearly
constant density throughout the region, totaling to several
solar masses of iron in a few parsecs (or more).
In this case the absorption feature can be caused by neutral iron 
and there is no problem in having the observed optical depth.
To make the feature disappear after a few seconds is highly problematic in
this scenario.
There are simply not enough photons to completely photoionize 
the iron, which therefore will absorb photons during the entire burst 
duration.
\vskip 0.3 true cm

\noindent
{\it Edge due to supernova ejecta ---}
If a SN exploded between a few months and a few years before 
the burst, then large amounts of iron--enriched material surrounding 
the burst are expected.
The total iron mass in this case is unlikely to exceed some fraction
of a solar mass, and the size of the region is of the order
of $10^{16}$--$10^{17}$ cm, depending on the SN--burst delay 
and on the velocity of the ejecta.
In this case if the disappearance of the absorption feature is due
to complete photoionization of iron,
then we require a large iron mass (tens of solar masses), unless
the absorbing iron is clumped (covering factor of few per cent)
and a clump lies by chance along the line of sight (see also
B\"ottcher et al. 2001).
\vskip 0.3 true cm

To decrease the iron requirement, one can envisage a situation in which
some recombination occurs, therefore allowing each iron atom to absorb
more than 26 photons.
This requires a higher density, since iron must recombine in
a timescale comparable to the ionization timescale; yet the plasma 
must remain thin to Thomson scattering.
Only in some very {\it ad hoc} geometries these two contradicting requirements
can be accommodated (e.g. in very narrow layers of material).

\vskip 0.3 true cm

\noindent
{\it Resonant line(s) due to supernova ejecta ---}
The large ionizing flux of the burst can lead, in a
short time, to complete or quasi--complete photoionization of
iron in the vicinity of the burst itself.
Because of this, and on account of the difficulties of the 
scenarios above, we investigated whether the absorption feature can
be an absorption line rather than an edge.
Since resonant lines are very narrow, this interpretation
requires a large spread of velocities, in order
to make the line detectable by the WFC.
In this scenario the disappearance of the feature 13 seconds after the burst
results from electron heating due to the illuminating flux.
Correspondingly the recombination rate decreases sharply.
What we find is intriguing.
First, the above argument allows to fix the distance of the
absorption material.
Second, assuming iron as the main absorbing agent,
we require velocities around $0.2c$. 
If this is the radial velocity of a SN ejecta,
we immediately derive that the SN exploded 100 days before the GRB.
This time implies that $^{56}Co$ is still abundant (as much as iron in fact)
and therefore that the absorption by $^{56}Co$ should also be present.
The energy resolution of the WFC cannot resolve the two lines, which
are therefore blended.
A consistent solution is found for $v\sim 0.13 c$.

\vskip 0.3 true cm

\noindent
{\it Small and very dense blobs ---}
In principle, absorption can be caused by small and dense 
blobs, very close ($10^{14}$ cm) to the burst site, therefore avoiding 
the requirement of a SN explosion before the burst. The density 
of the blobs can be high enough (and their temperature low enough) 
to ensure fast recombination. These blobs must be Compton thin
in order not to smear flickering in the light curve.
This leads to blobs with a typical size of the order of 100 km and densities
of the order of $10^{17}$ cm$^{-3}$, i.e. rather extreme values.
In addition, these blobs would be swept up by the fireball in
a short time, for typical values of the fireball Lorentz factor
(see Sect.~\ref{sect:vicine}).

\section{Neutral iron}
\label{sec:fei}

The straightforward interpretation of the feature reported by A2000
is that neutral iron is present along the line of sight to the burst. 
This iron is photoionized by the burst X--ray photons,
until all the electrons are stripped from the iron ions, causing
the disappearance of the feature. In order to check this simple
interpretation we must keep in mind the following three observational data:
(i) the optical depth of the feature was $\tau_{\rm edge} = 1.4$ (A2000); 
(ii) the time during which the edge was observed $t_{\rm edge} = 13$~s 
(A2000); 
(iii) the total
number of absorbed photons\footnote{The cosmological parameters 
$H_0 = 65$~km~s$^{-1}$~Mpc$^{-1}$ and $q_0=0.5$
are used to derive the value of $N_\gamma$.} 
$N_\gamma \simeq 3\times10^{57} \, f$,
where $f$ is the covering factor of the absorbing material surrounding
the burst\footnote{Note that a beamed fireball would introduce
an additional geometrical factor. 
This would reduce the number of photons and hence the mass
of iron required to produce the absorption feature. However, to use
this geometric factor to reduce the iron mass, a model must explain why 
iron is present only within the geometric cone of the fireball. 
In this paper we consider this possibility unlikely and we assume 
a spherical fireball for simplicity.}.
These three quantities are simply related to the geometrical distribution
of the absorbing material around the burst.
The edge opacity of a distribution of absorbers $n_{\rm FeI}(r)$ 
of neutral iron atoms is given by:
\begin{eqnarray}
\tau_{\rm edge} &=& \sigma_{\rm FeI} \, \int n_{\rm FeI}(r)\,dr 
\sim \nonumber \\
&\sim& n_{\rm FeI} \, \sigma_{\rm FeI}\,\Delta R =
{{M_{\rm FeI}\,\sigma_{\rm FeI}}\over{56\,m_p\,4\pi\,R^2}}
{{1}\over{f}},
\label{eq:taufei}
\end{eqnarray}
where 
$\sigma_{\rm FeI} = 1.2\times10^{-20}$~cm$^2$ is the photoionization 
cross section 
of K--shell electrons of neutral iron. The second line of Eq.~\ref{eq:taufei}
holds for a uniform density distribution $n_{\rm FeI}$ of iron atoms 
with a characteristic radius $R$ and width $\Delta R \ll R$.
The time after which the absorption feature disappears due to complete 
ionization of iron is given by:
\begin{equation}
t_{\rm edge} = 26\,Y\, \, \left[\int_{\nu_0}^\infty {{F(\nu)}\over{h\,\nu}}
\,\sigma_{\rm FeI}\, \left({{\nu}\over{\nu_0}}\right)^{-3}\,d\nu\right]^{-1},
\label{eq:tedg}
\end{equation}
where $F(\nu)$ is the flux density at the radius $R$.
Under the simplified conditions above, Eq.~\ref{eq:tedg} gives:
\begin{equation}
t_{\rm edge} = 26\,Y\,{{4\pi\,R^2\,\epsilon_{\rm ion}}\over
{L_{\rm ion}\,\sigma_{\rm FeI}}},
\label{eq:tedge}
\end{equation}
where $\epsilon_{\rm ion} \sim 10$~keV is the typical energy of the ionizing 
photons and $L_{\rm ion}$ the ionizing luminosity weighted over frequency
with the photoionization cross section.
The parameter $Y$, always smaller than unity, 
accounts for the fraction of electrons that are stripped from 
the iron atom as a consequence of the absorption of a photon 
(some of the electrons are ejected by the Auger effect). 
In addition, $Y$ accounts for the fact that the last two 
electrons contribute to absorption at different energies. 
The average value of $Y$ is in this case $Y=0.56$.

Equations~\ref{eq:tedge}, by itself, gives the characteristic distance
at which the absorbing iron must be located:
\begin{equation}
R = \left( {{L_{\rm ion}\,t_{\rm edge}\,\sigma_{\rm FeI}}\over
{26\,Y\,4\pi\,\epsilon_{\rm ion}}} \right)^{1/2} =
7 \times10^{17}\;{\rm cm}.
\label{eq:radius}
\end{equation}
If the iron is closer, the edge would live less than the 
observed 13 seconds. If, on the contrary, the absorbing iron is
more distant, an alternative mechanism to complete 
ionization is required to quench absorption.

Combining Eq.~\ref{eq:radius} with Eq.~\ref{eq:taufei}, the total mass of
iron can now be written:
\begin{eqnarray}
M_{\rm FeI} &=& {{56\,m_p}\over{26\,Y}}\,{{\tau_{\rm edge}\,L_{\rm ion}
\,t_{\rm edge}}\over{\epsilon_{\rm ion}}}\, f =
\nonumber \\
&=& 35\,f\,\left({{\tau_{\rm edge}}\over{1.4}}\right)\,
\left({{L_{\rm ion}}\over{10^{49} {{\rm erg}\over{\rm s}}}}\right)\,
\left({{t_{\rm edge}}\over{13{\rm s}}}\right)\, M_\odot.
\label{eq:mfei}
\end{eqnarray}
Note that all quantities in Eq.~\ref{eq:mfei} are measured, with the
exception of the covering factor $f$. Similar results are found, 
with detailed numerical computations, by B\"ottcher et al. (2001).

A consistency check can be made by considering constraint (iii). For a given 
distribution of absorbers, the number of absorbed photons is given by:
\begin{equation}
N_\gamma = 4\pi\,R^2\, t_{\rm edge}\,
\int_{\nu_0}^{\infty} {{F(\nu)}\over{h\,\nu}} \,
\left(1-e^{-n_{\rm FeI}\, \sigma_{\rm FeI}\, \left({{\nu}\over{\nu_0}}
\right)^{-3}\, \Delta R}\right)\, d\nu.
\end{equation}
Assuming, again, a uniform matter distribution,
this photon number is related to the total iron mass through:
\begin{equation}
N_\gamma = 26\,Y\,{{M_{\rm FeI}}\over{56\,m_p}} = 
{{\tau_{\rm edge}\,L_{\rm ion}\,t_{\rm edge}}\over{\epsilon_{\rm ion}}}\,f
= 1.4\times10^{58},
\label{eq:ngammas}
\end{equation}
which is a factor of $\sim 4$ larger than the (measured) value
given above, independent of the value of $f$.
Given the simplified geometry that is assumed, we do not consider 
this a major inconsistency.

An alternative possibility is to allow for recombination
of free electrons onto iron ions. 
In this case, each iron ion can absorb many more than $26\,Y$ photons,
thus decreasing the iron mass requirement.
In order for recombination to be efficient, its timescale
must be smaller than the ionization time, determined by the strong
ionizing flux.
The recombination time onto an ion with charge $Z$
is given by the Seaton (1959) formula which, interpolated
in the temperature range $10^2 < T_e < 10^6$~K can be expressed as:
$t_{\rm rec} \sim 4\times10^9\, T_e^{3/5}\, Z^{-2}\, n_e^{-1}$~s,
where $n_e$ is the electron density and $Z$ the ion charge.
The ratio of recombination to ionization time is then
given by:
\begin{equation}
{{t_{\rm rec}}\over{t_{\rm ion}}} = {{4\times10^9}\over{4\pi\,Z^2}}\,
{{T_e^{3/5}\,L_{\rm ion}\,\sigma_{\rm FeXXVI}}\over
{n_e\,R^2\,\epsilon_{\rm ion}}} \sim {{5\times10^{47} \, T_4^{3/5}}\over
{n_e\,Z^2\,R^2}} \; {\rm cm}^{-1},
\label{eq:trecion}
\end{equation}
where $\sigma_{\rm FeXXVI}$ is the photoionization cross section of 
H--like iron and $T_4=T_e/(10^4\, {\rm K})$. 
In order to keep iron in a low--intermediate ionization state 
(say $Z = 13$), the ratio of timescales in Eq.~\ref{eq:trecion} 
must be $\lesssim 1$, implying:
\begin{eqnarray}
n_e\,R^2 \left({{Z}\over{13}}\right)^2 \, T_4^{-3/5}
\gtrsim 3 \times 10^{45} \; {\rm cm}
&\Rightarrow& \nonumber \\
\Rightarrow \tau_T\,{{R^2}\over{\Delta R}}\,
\left({{Z}\over{13}}\right)^2 \, T_4^{-3/5}
\gtrsim 2\times10^{21} && \; {\rm cm},
\label{eq:cond}
\end{eqnarray}
where $\tau_T$ is the Thomson optical depth of the absorbing material.
Since the Thomson depth along the line of sight cannot be larger than
unity (otherwise the flickering behavior of the burst light curve
would be smeared out) Eq.~\ref{eq:cond} can be rewritten as: 
\begin{equation}
{{\Delta R}\over{R}} \lesssim {{R\,\left({{Z}\over{13}}\right)^2\, T_4^{-3/5} }
\over{2\times 10^{21} \,{\rm cm}}}.
\label{eq:cond2}
\end{equation}
In addition, we must consider that recombination can be important 
to reduce the total iron mass only if its timescale is in the order 
of seconds. This requires, with a derivation analogous to that 
of Eq.~\ref{eq:cond2}:
\begin{equation}
\Delta R \lesssim 2.5 \times 10^{14} \left({{Z}\over{13}}\right)^2
\, T_4^{-3/5} \; {\rm cm}.
\label{eq:cond3}
\end{equation}
Equations~\ref{eq:cond2} and~\ref{eq:cond3}
show that, to have recombination in a Thomson thin 
medium, the geometrical depth along the line of sight must be many orders 
of magnitude smaller than the distance from the bursting source. 
We conclude that such geometry is extreme.
A blobby medium might instead be a more realistic 
geometrical setup.

\section{Hydrogen--like iron}
\label{sec:fex}

Equation~\ref{eq:cond2} shows that during the prompt emission it is highly 
unlikely that iron atoms in low--intermediate ionization states can survive
in the surroundings of the GRB. On the other hand, the feature
detected by A2000 has a rest frame frequency consistent with the K--shell 
photoionization edge of neutral iron. 
How can this riddle be solved?
We consider in this section the possibility that the iron
along the line of sight to the burster is highly ionized. 
Given the results above, we are
justified in considering only two species of iron ions: fully ionized iron 
(FeXXVII) and H--like iron (FeXXVI). 
This is a reasonable approximation since the ionization
timescale of FeXXVI to FeXXVII is very short and 
the probability to have a second electron recombining in such a 
short timescale is small.
Even though the vast majority of iron ions is expected to be 
fully ionized, recombination plays a crucial role in reducing the total
amount of iron required to reproduce the observed feature. 
In fact, as described below, each iron ion may recombine $\sim 1000$ times
during the first 13 seconds of the burst, absorbing many more than the
$26\,Y$ photons discussed in Sect.~\ref{sec:fei}. 
On the other hand, the disappearance of the feature 13 seconds after the burst 
onset requires that recombination is halted after that time, making
the iron opacity decrease drastically.
There are in principle two ways to achieve this: in fact recombination
becomes slower if the electron density decreases and/or the electron 
temperature increases. 
In the first case, we have $d\tau / \tau = -3\,dr/r$: a sizable reduction 
of the opacity can be achieved if the radius of the absorbing shell is 
increased by one third in the $\sim 13$ seconds during which the edge 
is seen.

However, this would require relativistic speeds
(even for the smallest radii allowed by the observations), and
the edge would then be Doppler shifted to a very different energy.

Consider instead the effect of the illumination of the material by
the burst photons: this will increase the electron temperature
after a few scatterings, i.e. on a timescale:
\begin{equation}
t_{\rm T} = { 4\pi\,R^2\,\epsilon \over L\,\sigma_{\rm T}},
\end{equation}
where $L$ is the isotropic equivalent luminosity of the burst and 
$\epsilon \sim 500$~keV the typical photon energy\footnote{
Note that we are here neglecting Klein Nishina effects for simplicity.}.
Setting the heating timescale equal to the lifetime of the edge 
($\sim 10$~s) we obtain:
\begin{equation}
R = \left( {{10\,L\,\sigma_T}\over{4\pi\,\epsilon}} \right)^{1/2}
= 2.6\times10^{16} \, \left({{L}\over{10^{51}\,{{\rm erg}\over{s}}}}
\right)^{1/2} \;\; {\rm cm}.
\label{eq:rad}
\end{equation}
Note that this estimate is independent of the electron density, as long as
each electron can scatter more than one incoming photon in the burst 
duration time, i.e. as long as $t_T$ is shorter than the burst duration.
Moreover, the accelerated electrons share their energy with the 
plasma on a timescale set by the ``slowing--down'' time (Spitzer
1956, eq. 5-28) which is of the order of a hundred seconds in the 
physical conditions described above ($\gamma_e \sim 2$,
$T_e\sim 10^4$~K, $n_e\sim 10^{11}$~cm$^{-3}$). 
We conclude that the requirement 
on the timescale of electron heating is thus a powerful way to 
constrain the radius of the absorbing plasma.

\subsection{Resonant scattering on FeXXVI}
\label{sec:reso}

A detailed treatment of the resonant scattering of photons from
H--like iron is beyond the scope of this paper and we
refer the reader to the classical text of Rybicki and Lightman (1979)
and the theoretical paper of Matt (1994).
We consider here the resonant scattering on H--like iron, 
and in particular the resonant transition  $1s-2p$ (Matt 1994), 
which has an oscillator strength $f_{lu}= 0.416$ and a rest 
frame frequency $h\nu = 6.927$~keV (Kato 1976).

We investigate the equivalent width (EW) and the depth
of the feature produced by the resonant scattering from FeXXVI. 
The feature produced by a single iron atom is deep 
(the resonant cross section at the core of the feature 
is $\sigma_{1s-2p} \approx 2 \times 10^{-16}$~cm$^{2}$) and narrow
($\Delta \epsilon \sim 3.5$~eV) and for this reason 
an equivalent width of $\sim 1$~keV cannot be obtained, unless
the resonance is broadened by intrinsic velocity
dispersion of the absorbing material.

Given the properties of the feature detected in GRB~990705,
absorption is in the optically thick regime, and the 
EW of the resonance feature cannot be computed 
analytically (Matt 1994).
We hence computed numerically the opacity of the feature
as a function of the frequency convolving the single atom 
Lorenzian profile (Rybicky and Lightman 1976) with an appropriate
velocity distribution. 
To model an outflow with a one parameter velocity distribution, 
we adopted a Maxwellian function of the form:
\begin{equation}
p(v) = {{4\,v^2}\over{\sqrt{\pi}\,v_0^3}}\,e^{-\left({{v}\over 
{v_0}}\right)^2} \qquad \;\; v>0,
\end{equation}
which is characterized by a mean velocity $\langle v \rangle \sim v_0$
and FWHM $\sim v_0$. Since this is a directed outflow velocity, the 
centroid of the absorption is blueshifted. This is taken into account
when the absorption profile is modelled.
We considered also different velocity distributions. However,
after convolution with the instrumental resolution of the WFC,
the functional shape of the velocity distribution does not play 
any relevant role. Since the scenario involves a supernova remnant (SNR)
(see Sect.~\ref{sec:gg}), we consider here that the iron ions are likely
to be the result of the decay chain 
$^{56}$Ni~$\rightarrow\,^{56}$Co~$\rightarrow\,^{56}$Fe.
Fixing the remnant radius to the value derived in Eq.~\ref{eq:rad},
we obtain the age of the SNR as a function of
the velocity $v_0$ and consistently compute the fraction of 
$^{56}$Co still present (see Vietri et al. 2001). 
The fraction of iron ions is shown in the upper panel
of Fig.~\ref{fig1} as a function of the expansion velocity $v_0$.
In some peculiar SNe, the iron may
be directly injected in the remnant by the SN explosion if part of the
$^{56}$Fe of the nucleus of the progenitor is expelled or, alternatively,
if neutronization in the innermost layers of the (ejected) mantle is high 
enough that $^{54}$Fe (rather than $^{56}$Ni) is preferentially 
synthesized (see, e.g., Limongi et al., 2000 and references therein).
The results of this paper
still hold, provided that the column density 
$N_{\rm FeXXVI+CoXXVII}$ is substituted with $N_{\rm FeXXVI}$
and that slightly larger velocities $v_0$ are considered.
This happens because the different energy of the Fe and Co 
transitions acts like an additional velocity dispersion.

In addition to the resonance feature, we consider the effect of
photoionization of the same K--shell electrons which produce
the resonant transition above. Note that for FeXXVI the
photoionization threshold is at a frequency $h\nu = 9.28$~keV
(observed $h\nu \sim 5$~keV), well outside the energy range
in which absorption is observed in GRB~990705
(see Fig.~\ref{fig3}). Again we consider
also photoionization of cobalt ions.

The result of the numerical computation of the EW of the resonance
and photoionization features is plotted in Fig.~\ref{fig1}
as a function of the outflow velocity parameter $v_0$,
ranging from $10^9$ to $10^{10}$~cm~s$^{-1}$. Five different
values of the FeXXVI+CoXXVII column density are considered
(see caption).
The photoionization EW (dashed lines) is less sensitive to the
velocity dispersion, since the feature is naturally broad and 
is not in the optically thick regime even for the
smallest velocity dispersion. The resonance feature, on the other hand,
reaches a $\sim 1$~keV EW only for velocities larger than $0.1\,c$.
In all cases, the photoionization EW is larger than the resonance EW.

Despite this, in many conditions the resonance feature is much deeper
than the photoionization feature, since the width of the latter
is much broader than the instrumental resolution.
In Fig.~\ref{fig2} we show a collection of synthetic spectra
computed for $N_{\rm FeXXVI+CoXXVII} = 6\times10^{19}$~cm$^{-2}$
and for velocities ranging from $10^9$ to $8\times 10^{9}$~cm~s$^{-1}$.
A power--law continuum with photon index $\Gamma_{\rm ph} = -1.1$ 
is assumed.
The dashed lines show the intrinsic spectra, while the solid lines
show the spectra as observed with the WFC instrumental resolution.

For small values of the velocity dispersion $v_0$, the depth of the 
feature observed with the WFC is very small (a very deep and narrow
feature is smoothed out by the instrument response).
For higher values of both
the velocity dispersion and the column density,
the feature can be very deep, reaching a core optical depth 
$\tau \sim 1.5$ (last three spectra of Fig.~\ref{fig2}).

In order to carry out a detailed comparison with the observations 
of GRB~990705, we included this model in XSPEC (Arnaud 1996) as a 
tabulated multiplicative model. To fit the spectrum, an absorbing 
column density in addition to the Galactic $N_H$ has been included
({\it ZWABS} model), with the redshift of the burst. 
This absorber has solar abundances
of elements, and is considered to be more distant from the burst
with respect to the resonant iron. In fact, at the distance of the
resonant iron, the ionization parameter is so high that all
lighter elements are completely ionized and photoionization
absorption is negligible.

The best fit ($\chi^2/d.o.f. = 5.1/7$)
is obtained for $v_0 = 4.0^{+3.0}_{-2.5}\times10^9$~cm~s$^{-1}$ and 
$N_{\rm FeXXVI+CoXXVII} = 7.0^{+3.0}_{-4.5}\times10^{19}$~cm$^{-2}$,
where the uncertainties give the single parameter errors at the
$90\%$ confidence level.
The best fit model is plotted,
overlaid on the deconvolved data, in Fig.~\ref{fig3},
while Fig.~\ref{fig4} shows the confidence region for the 
two parameters at the $1\sigma$, 90 and $99 \%$  confidence
level.

\subsection{Physical conditions of the absorbing material}

We derived from the above analysis that a deep resonant feature
accompanied by a moderate photoionization feature can be produced by
a cloud fulfilling these two conditions:
i) the column density of H--like iron
$N_{\rm FeXXVI}$ is $\approx 4\div 10 \times 10^{19}$~cm$^{-2}$;
ii) the velocity $v_0$ is $\sim 0.1\,c$ with comparable velocity dispersion.

We now analyze whether these conditions are likely to be present
in the surroundings of GRBs. 
First we require, as discussed in Sect.~\ref{sec:fei}, that the 
Thomson opacity of the absorbing medium is smaller that unity. 
We obtain:
\begin{equation}
\tau_T = {{N_{\rm Fe}\,\sigma_T}\over{4.68\times10^{-5}\,A_{\rm Fe}}}
= {{0.9}\over{\eta\,A_{\rm Fe}}}\,\left({{N_{\rm FeXXVI}}\over{
7\times10^{19}\,{\rm cm}^{-2}}}\right),
\label{eq:taut}
\end{equation}
where $A_{\rm Fe}$ is the iron abundance in solar units 
(Anders \& Grevesse, 1989) and the parameter
$\eta = N_{\rm FeXXVI}/N_{\rm Fe}$ gives the ratio of iron XXVI to total iron. 
Since $\eta$ has likely a value around a few per cent,
Eq.~\ref{eq:taut} implies that the absorbing material is overabundant in iron
with respect to the solar value. 
Note that iron enrichment, on a similar basis, is required also in 
the case of the photoionization of neutral iron discussed 
in Sect.~\ref{sec:fei}. 
In the case of Sect.~\ref{sec:fei},
$\tau_T = 1.8\, \tau_{\rm FeI} / A_{\rm Fe} \sim 2.5/A_{\rm Fe}$.
A2000 derive $A_{\rm Fe} = 75$. 
This value is obtained 
by comparing the depth of the feature with the low energy ($h\nu < 2$~keV)
absorption due to the photoionization of lower $Z$ elements. 
This comparison, however, gives the ratio of abundances of ions 
that can absorb photons, i.e. of ions with at least one electron.
Since the ionization state of elements with $Z<26$ is higher
(if not complete) than the ionization state of iron, the iron abundance 
is overestimated by this method.
For this reason we adopt here a conservative fiducial
value of $A_{\rm Fe} = 10$.

Rewriting Eq.~\ref{eq:taut} in terms of the population ratio $\eta$
and using our fiducial value for the iron richness
we obtain $\eta \gtrsim 0.1$. 
Consider a two level system of atoms, with 
$t_e$ and $t_d$ the times of excitation and decay, respectively.
The fraction of atoms in the ground level is given by
$f_d = t_e/(t_e+t_d)$ which, in the limit $t_e\ll t_d$
can be approximated as: $f_e=t_e/t_d$.
The same conditions apply to the absorbing iron ions, yielding:
\begin{equation}
\eta \approx {{t_{\rm ion}}\over{t_{\rm rec}}} = 
{{4\pi\,R^2\,\epsilon_{\rm ion}} \over{L_{\rm ion}\,\sigma_{\rm FeXXVI}}}\,
{{n_e}\over{6\times10^6\,T_e^{3/5}}},
\end{equation}
from which we can obtain the electron density $n_e$:
\begin{eqnarray}
n_e &=& {{L_{\rm ion}\,\sigma_{\rm FeXXVI}\,6\times10^6\,T_e^{5/3}}\over
{4\pi\,R^2\,\epsilon_{\rm ion}}}\, \eta = \nonumber \\
&=& 8.3\times10^{10} \, T_4^{3/5}\left({{1}\over{\tau_T}}\right)\,
\left({{10}\over{A_{\rm Fe}}}\right),
\end{eqnarray}
where we use the radius given by Eq.~\ref{eq:rad}.
Requiring again that the Thomson depth is less than unity, we obtain
the geometrical thickness of the absorbing material:
\begin{equation}
\Delta R = {{\tau_T}\over{n_e\,\sigma_T}} = 2\times10^{13}\,
\left({{\tau_T}\over{1}}\right)^2\, \left({{A_{\rm Fe}}\over{10}}\right)
\;{\rm cm}.
\end{equation}
This is much smaller than the distance $R$ of the absorbing
material from the burst, and for this reason a shell geometry is 
unlikely also in this case.
A more realistic scenario is a clumpy medium, with overdensities
of the size described above embedded in a lower density medium, 
with a density contrast of the order of $100\div1000$.
Note that the clumpiness discussed here is different from
that required to reduce the unrealistic iron mass of Eq.~\ref{eq:mfei}.
In fact, the covering factor of blobs can here be unity.
Moreover, Eq.~\ref{eq:ngammas} does not hold in this recombining clouds.
Given the required velocity dispersion, an even more likely scenario
is that the absorption feature is produced by the superposition
of many smaller blobs with different velocities.

If the covering factor of the clouds, and hence the column density
of iron, is reasonably uniform, the total iron mass surrounding the burst
location can be estimated:
\begin{equation}
M_{\rm Fe} = 4\pi\,R^2\,N_{\rm Fe}\, m_{\rm Fe} \sim 0.16 M_\odot,
\end{equation}
where $N_{\rm Fe} = N_{\rm FeXXVI}/\eta$ is the total iron column density.
This value is well within the range of iron mass produced by
SNe, and a factor of 200 smaller than the mass required
in models without recombination (see Eq.~\ref{eq:mfei}).

A velocity dispersion comparable to that required
to produce a deep feature was measured in the
emission line of GRB~991216. 
Piro et al. (2000) find $\delta v_{\rm FWHM} \sim 0.15\,c$.
If the iron in the surrounding of the burst is supplied by a
young SNR, as discussed also for emission
features of GRB afterglows (Lazzati et al. 1999, Vietri et al. 2001), 
then large velocity dispersions are naturally expected, 
due to the differential expansion of the remnant. 

\section{Geometry and radius}
\label{sect:vicine}

The detection of iron emission features in the X--ray afterglows 
of GRBs is considered among the most reliable signatures of the 
association of the bursters with the death of massive stars. 
It is generally believed that if GRBs are associated with the explosion
of a star, then the iron features are produced by the iron synthesized
during the explosion of the progenitor star
itself (Lazzati et al. 1999; see also B\"ottcher 2000).
The results in Sect.~\ref{sec:fei} and Sect.~\ref{sec:fex} clearly
point to iron at a radius $R>10^{16}$~cm from the burster.
This means that the burst and the iron could not be produced
by the same explosion, since the iron must be moving at sub--relativistic
speeds from the place of its synthesis to this radius before the GRB onset.
The absorbing iron is in fact detected at the rest frame frequency,
limiting its possible proper motion to sub--relativistic speeds.
Yet it is important to ascertain whether it is possible to produce a 
similar absorption feature from smaller radii such that the iron 
could have been ejected just a few hours before the burst (some iron 
may be ejected a few dynamical time--scales before the star 
collapse). 
This implies that the absorbing iron must be located at a radius 
$R \sim v_{\rm Fe}\,t_{\rm dyn} \lesssim 10^{14}$~cm. 

Let us consider absorbing iron located at such a small distance.
A first requirement is that the fireball reaches this
distance in an observed timescale longer than 10 seconds, to avoid
sweeping up the absorbing material, quenching its absorbing power or, at 
least, boosting it to highly blueshifted frequencies. 
The radius of the fireball after 10 seconds is 
$R_{\rm FB} \sim 3\times 10^{11} \, \Gamma^2$.
This implies $\Gamma \lesssim 20$, a very small value compared to
$\Gamma \sim 100\div300$ which is required in the fireball model to avoid 
the so--called compactness problem (see, e.g., Piran 1999).
However it is possible that, since high ambient densities are in any case
required, the fireball enters the slowing--down afterglow phase at
smaller radii. Therefore we do not consider this a major problem.

Consider now the amount of iron required. If recombination does not play 
an important role, the largest column density of iron can be achieved in the 
extreme conditions of a cloud of pure iron (the Thomson depth of the 
absorbing medium must be less than unity). In this case the iron 
column density is bound to be less than 
$N_{\rm Fe} \le (26\,\sigma_T)^{-1} = 5.8 \times 10^{22}$~cm$^{-2}$.
In the absence of recombination, the total number of photons that 
this iron can absorb through photoionization is:
\begin{equation}
N_\gamma = {{4\pi\,R^2}\over{\sigma_T}}\, Y \sim 2\times10^{53}\,Y,
\end{equation}
where $Y$ is defined in Eq.~\ref{eq:tedg}. The numerical value above is
four orders of magnitude smaller than the inferred number of absorbed photons
in GRB~990705.

If, on the other hand, recombination is efficient, Eq.~\ref{eq:cond2}
can be rewritten as  $\Delta R/R < 5\times10^{-8} \,(Z/13)^2$. 
An extremely thin shell or, alternatively, an extreme level of
clumpiness would be required.
The absorbing iron should be contained in blobs with radius
$R_{b} \sim 10^{7}$~cm, with a particle density 
$n_{\rm b} \sim 10^{17}$~cm$^{-3}$. The issue is whether these 
blobs can survive in pressure equilibrium in the ambient medium.
Since Thomson thinness must hold for the ambient
medium as well as the blobs, pressure equilibrium implies
$T_{\rm b}\, n_{\rm b} = T_{\rm a}\, n_{\rm a}$, where the subscripts
b and a refer to the blobs and ambient medium,
respectively. If, as assumed in deriving Eq.~\ref{eq:cond2}, 
$T_{\rm b} \sim 10^4$~K, then $T_{\rm a}$ is expected in the $10^{11}$~K
range, a very high value.
Additional effects, such as magnetic confinement of the 
blobs, would hence be required. We regard this scenario as unlikely.

Yet, a significant degree of clumpiness of the absorbing medium 
appears to be required in all the scenarios discussed in this
paper. Absorption from neutral iron requires
clumpiness in order to reduce the covering factor of the absorbing material
and consequently the total iron mass. Resonant scattering from FeXXVI
requires moderate clumpiness (density contrast of $\sim 1000$)
in order to allow for electron recombination on FeXXVII ions.

\section{Summary and discussion}

We investigated in this paper the implications of an absorption
feature in the prompt X--ray spectrum of a GRB, based on the properties 
measured with the {\it Beppo}SAX WFC in GRB~990705. 
We find that the only scenario which can naturally explain the observed 
properties of the absorption feature  of GRB~990705 is the explosion of 
the burst within a young SNR (see e.g. the
Supranova scenario of Vietri and Stella, 1998). 
This can produce the large 
amount of iron--enriched material required to absorb
X--ray photons by resonant scattering
and naturally account for the expansion velocity required to 
broaden the resonance feature.
In this scenario, burst photons are produced by internal shocks
at the internal shock (or transparency) radius, $\sim 10^{13}$~cm,
and propagate unaffected until they reach a dense region
(the SNR) with high iron abundance.
X--ray photons are scattered away from the line of sight by resonant 
scattering in the $1s$--$2p$ transition of FeXXVI and CoXXVII, 
while photons outside of the resonance are preferentially 
scattered by free electrons. Before free electrons are 
heated by GRB photons, the fast recombination of 
``cold'' electrons allows each iron ion to absorb more than 100
photons and therefore give rise to a large opacity.
After several seconds, however, the free electrons are heated at the 
Compton temperature of the burst photons and recombination is inhibited,
so that the iron opacity becomes negligible. The radius of the SN
shell is fixed by the requirement that the heating timescale is several
seconds ($R\sim 2\times 10^{16}$~cm). A uniform density shell at this 
radius, however, would not be dense enough to have a sufficiently
fast recombination of iron ions. In order to reproduce the observed 
absorption the remnant matter must be clumpy, with a density contrast of 
$\sim 1000$.

In this scenario photons are scattered away from the line of sight.
An iron emission line, however, is not observed in the spectrum 
of GRB~990705. 
Two mechanisms reduce the flux of the emission line. 
In the first $t_{\rm edge} \sim 10$~s of the burst, 
the line photons are diluted by geometrical effects and the
luminosity of the line is multiplied by a factor 
$c\,t_{\rm edge} / R \sim 10^{-5}$. 
Moreover, if the fireball is beamed in a cone, absorption 
is directional while re--emission is isotropic. 
This reduces again the line flux by an extra factor $\Omega / 4\pi$.
These effects explain the non detection of an emission line
during the WFC observation of GRB~990705. 
The predicted luminosity of the emission line is 
$L_{K\alpha} = N_\gamma \, \epsilon_{K\alpha}\, c / 
R \,\, (\Omega/4\pi) \sim 4.8\times 10^{43}\, (\Omega/4\pi)$~erg~s$^{-1}$ 
which, for a cosmological burst at $z\sim1$ gives a flux
$F_{K\alpha} \sim 1.5 \times 10^{-14}\, (\Omega/4\pi)$~erg~cm$^{-2}$~s$^{-1}$. 
This flux, undetectable during the burst, is expected to lasts for
$t_{K\alpha} \sim R/c \sim 10$~days. 
Such a line intensity, when the continuum has faded, should be easily
detectable with the current generation X--ray satellites such as
Chandra and XMM--Newton and is a powerful tool to constrain the beaming 
geometry of the fireball.

The SNR is reached by the fireball 
$R/(c\,\Gamma^2) \sim 60 (\Gamma/100)^2$~s after the burst explosion and
the fireball itself is slowed down to sub relativistic speeds as a 
consequence of the large mass swept up in the impact
(Vietri et al., 1999). For this reason, if the absorption feature
is real, an usual power--law afterglow lasting for months
can not be associated with GRB~990705.
Unfortunately both optical and X--ray observations
of the afterglow are not conclusive. 
In X--rays, a fading source was detected within the burst error box, 
but the statistics is not sufficient to draw any detailed conclusion 
on the decaying law (A2000).
In the optical and near infrared the source was detected only once in 
the $V$ band and twice in the $H$ band, from few hours to $\sim$~one day
after the GRB trigger (Masetti et al., 2000). 
The first two $H$ band measurements define a power--law decay 
with index $-1.4$, but a third attempt to detect the source gave an upper 
limit much dimmer the predicted power--law decline.
If, on the other hand, the X--ray/optical
transient associated to the burst is due to the deep impact of the fireball
on the remnant (Vietri et al., 1999), the radiation is isotropic and 
the flux should be constant over a timescale $R/c \sim 10$~days.
The reality lies probably in the middle: what we may have observed
is a regular power--low, running into a non--relativistic Sedov--Taylor
solution in a shorter time than in other GRBs due to the very high 
density of the 
remnant. The timescale of this transition is in fact approximately given
by the width of the remnant over the speed of light, i.e. one day.

The fact that the edge has been observed in a burst with such a peculiar 
afterglow may well not be a coincidence.

\begin{acknowledgements}
We thank the referee, Markus B\"ottcher, for his constructive and careful
comments. We thank Martin Rees and Silvano Molendi for useful discussions and 
Jens Hjorth for informing us about the optical measurement of the 
redshift of GRB~990705 prior to publication. This work was partially
supported through COFIN grants.
\end{acknowledgements}

\clearpage
\centerline{\psfig{file=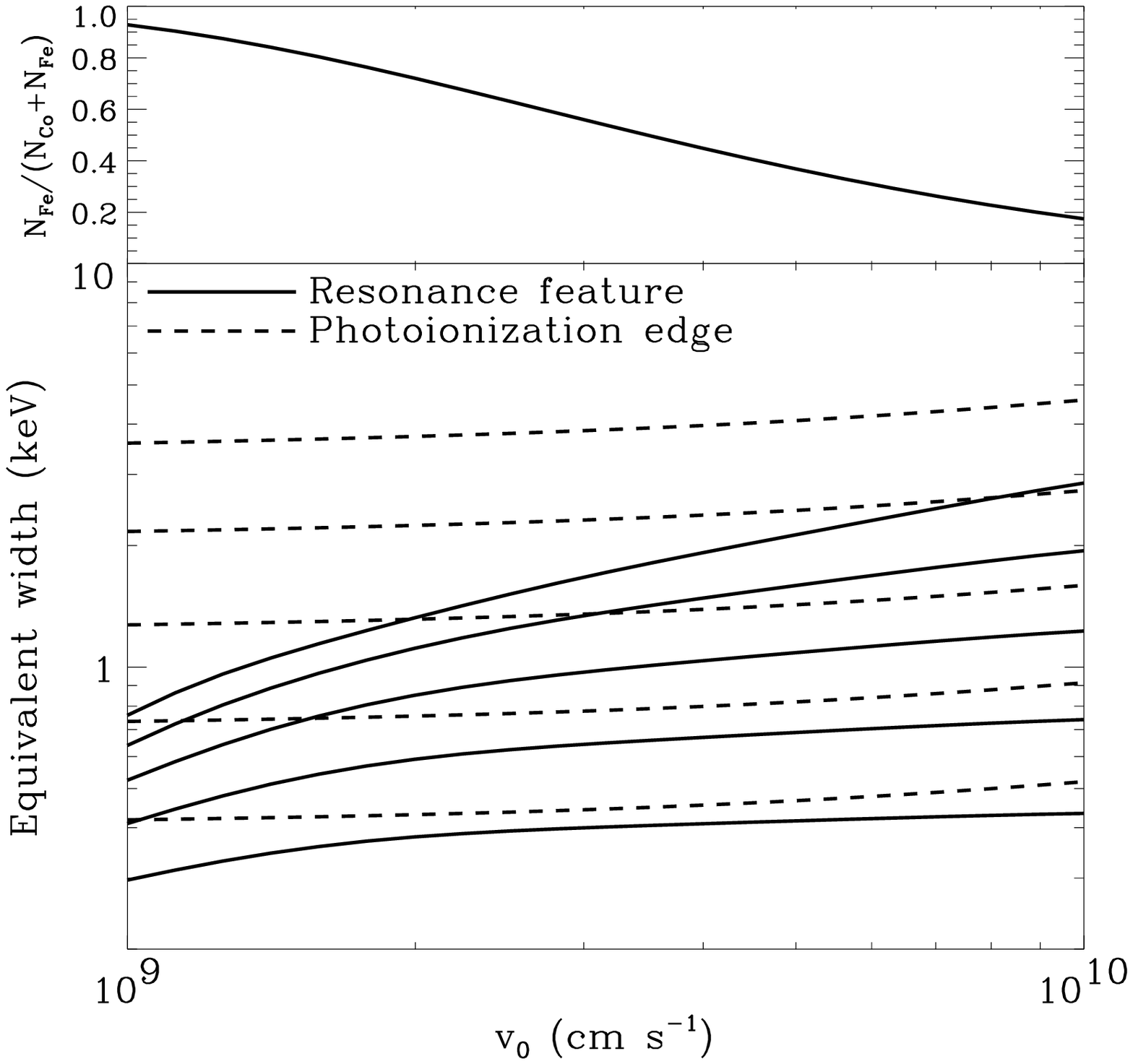}}
\figcaption{{Equivalent width of the resonance and photoionization 
features (lower panel) as a function of the velocity of the SNR (see text). 
Dashed lines show the EW of the photoionization feature, while
solid lines show the EW of the resonance feature. From bottom to top,
five values of the column density are plotted:
$N_{\rm FeXXVI+CoXXVII} = 10^{19}$, $1.78\times10^{19}$, 
$3.16\times10^{19}$, $5.26\times10^{19}$ and $10^{20}$~cm$^{-2}$.
The upper panel shows the ratio of iron ions over the sum of iron plus
cobalt ions. For larger velocities, the lifetime of the SNR is 
smaller and hence the fraction of iron smaller.}
\label{fig1}}

\clearpage
\centerline{\psfig{file=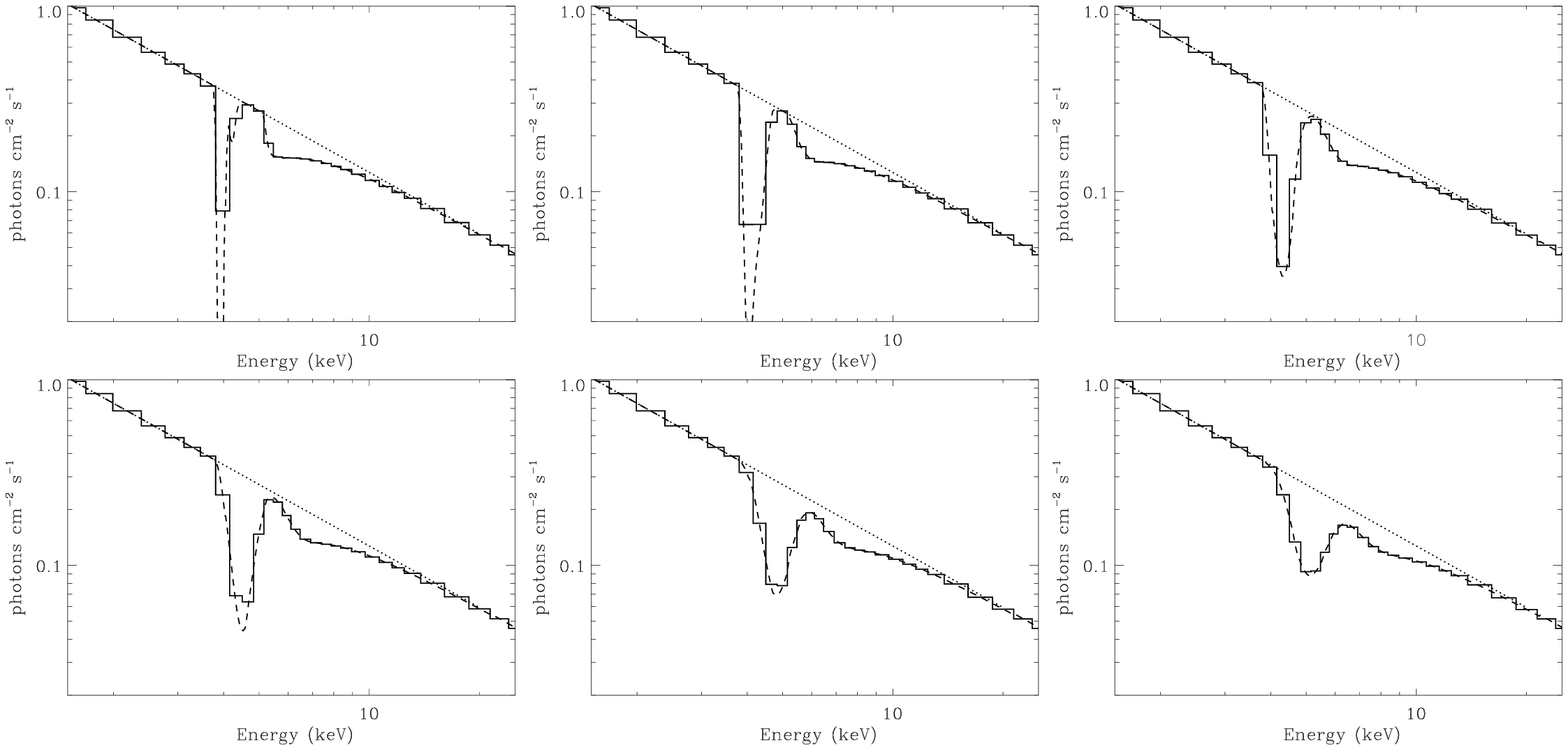,width=\textwidth,angle=90}}
\figcaption{{Synthetic spectra showing the impact of the resonant
and photoionization features for different outflow velocities.
The solid line show the spectra as observed after convolution
with the WFC instrumental resolution, while the dashed line shows
the intrinsic spectrum.
The column density of absorbers is set in all panels 
$N_{\rm FeXXVI+CoXXVII} =6\times10^{19}$~cm$^{-2}$. From left to right
and top to bottom, the six panels have outflow velocities
$v_0= 10^9$, $2\times10^{9}$, $3\times10^9$, $4\times10^9$,
$6\times 10^9$ and $8\times10^9$~cm~s$^{-1}$. The presence
of lower atomic number elements is neglected
in the computation of these spectra.}
\label{fig2}}

\clearpage
\centerline{\psfig{file=f3.eps,width=\textwidth,angle=-90}}
\figcaption{{Best fit model compared to the deconvolved
data from A2000. The best values of velocity dispersion and column density
are: $v_0 = 4 \times 10^9$~cm~s$^{-1}$ and 
$N_{\rm FeXXVI+CoXXVII} = 7\times10^{19}$~cm$^{-2}$. }
\label{fig3}}

\clearpage
\centerline{\psfig{file=f4.eps,width=\textwidth,angle=-90}}
\figcaption{{Confidence regions in the $v_0$-$N_{\rm FeXXVI+CoXXVII}$
plane. Contours show the $1\sigma$, $90$ and $99 \%$ confidence
level. The coutours are not closed in the upper
part of the figure (for high column densities). 
Column densities $N_{\rm FeXXVI+CoXXVII}>10^{20}$~cm$^{-2}$
are in any case phisically unlikely, because they would imply extremely iron
(cobalt) rich material.}
\label{fig4}}

\end{document}